\documentclass[aps,pra,twocolumn,amsmath,floatfix,showpacs,superscriptaddress]{revtex4}

\usepackage{graphicx}
\usepackage{color}
\usepackage{marvosym}
\usepackage{wasysym}
\usepackage{pifont}
\usepackage{amssymb}
\usepackage{amsmath}
\usepackage{amsfonts}
\usepackage{epsfig}
\usepackage[american]{babel}

\newcommand{\ud}{\mathrm{d}}
\newcommand{\vv}[1]{\tilde{#1}}
\newcommand{\ie}{{\rm i.e.}}
\newcommand{\eg}{{\rm e.g.}}
\newcommand{\br}{{\bf r}}

\begin{document}

\title{Splitting dynamics of giant vortices in dilute Bose-Einstein condensates}

\author{Pekko Kuopanportti}
\email{pekko.kuopanportti@tkk.fi}
\affiliation{Department of Applied Physics/COMP, Aalto University, P.O. Box 15100, FI-00076 AALTO, Finland}
\author{Mikko M\"ott\"onen}
\affiliation{Department of Applied Physics/COMP, Aalto University, P.O. Box 15100, FI-00076 AALTO, Finland}
\affiliation{Low Temperature Laboratory, Aalto University, P.O. Box 13500, FI-00076 AALTO, Finland}
\date{\today}

\begin{abstract}
We study the splitting of multiply quantized vortices with large quantum numbers in dilute nonrotated Bose-Einstein condensates in the zero-temperature limit. The splitting is observed to result in vortex-free condensate fragments which are separated by vortex sheets. The number of these fragments is found to be equal to the angular-momentum quantum number of the Bogoliubov excitation mode responsible for the splitting, although the formulation of the fragments cannot be described by small-amplitude excitations. Thus, the realization of an isolated giant vortex and the observation of its splitting would provide a means to directly relate the experimental data to discrete theoretical quantities.
\end{abstract}

\hspace{5mm}


\maketitle

\section{\label{sc:intro}Introduction}

The creation and observation of quantized vortices in dilute Bose-Einstein condensates (BECs) of alkali-metal atoms in 1999~\cite{Matthews1999} was an important demonstration of the superfluid properties of these systems. Since then, the study of vortices in BECs has flourished both theoretically and experimentally~\cite{Fetter2009}. 

When a large amount of angular momentum is transferred to a dilute BEC, it typically responds by nucleating vortices. If a condensate in a harmonic trap is subjected to rapid rotation, large arrays of singly quantized vortices are observed~\cite{Madison2001,Abo-Shaeer2001,Coddington2003,Bretin2004,Stock2005}. However, the condensate can also acquire high angular momentum through multiquantum vortices, for which the phase of the condensate order parameter winds an integer multiple $\kappa$ of $2\pi$. Multiquantum vortices have been created in dilute BECs by using a focused laser beam to remove atoms from the center of a rotating condensate~\cite{Engels2003,Tapsa2004,Tapsa2005}, by transferring angular momentum into the condensate by a Laguerre-Gaussian laser beam~\cite{Andersen2006,Tapsa2008}, and by a topological phase engineering method~\cite{Ville2008} which utilizes the spin degree of freedom of the condensate~\cite{Ueda2010} and its coupling to an external magnetic field~\cite{Nakahara2000,Isoshima2000,Ogawa2002,Mikko2002,Leanhardt2002,Leanhardt2003a,Kumakura2006}. 

Several theoretical studies~\cite{multi1,multi2,Jackson2005,multi3,multi4,Kuopanportti2010} have shown that states with a multiquantum vortex are typically dynamically unstable in harmonically trapped BECs. Dynamical instability is a peculiar feature of nonlinear dynamics, and in the case of multiquantum vortices, it causes the vortex to split into single-quantum vortices even in the absence of dissipation~\cite{splitting1,splitting2,splitting4,splitting5,splitting3,splitting6}. On the other hand, various studies have also addressed different means to stabilize multiquantum vortices, \eg, by rotating the condensate in the presence of a plug potential~\cite{Tapsa2002} or an anharmonic trapping potential~\cite{Lundh2002,Kasamatsu2002,Fischer2003,Josserand2004}. Stable multiquantum vortices have also been found to exist in two-component BECs~\cite{Ruostekoski2004,Christensson2008}.

Recently, it has been suggested that vortices with arbitrarily large winding numbers could be achieved in BECs with the so-called vortex pump, \ie, by cyclically pumping vorticity into the condensate~\cite{pumppu}. The technique is based on the topological phase engineering method: the spin degree of freedom of the condensate is controlled locally by adiabatically tuning external magnetic fields, which renders the system to acquire a fixed amount of vorticity in each pumping cycle. In the cycle presented in Ref.~\cite{pumppu}, a homogeneous magnetic field is used alongside alternating quadrupole and hexapole magnetic fields. With this setup, the vortex pump can be operated both fully adiabatically and partly nonadiabatically. In the former case, an additional optical potential is needed to confine the condensate during the cycle. Furthermore, in order to stabilize the vortex and prevent it from splitting prematurely, an optical plug potential can also be employed during the pumping stage. The plug can be realized by a focused laser beam as was done, e.g., in Ref.~\cite{Tapsa2005}. The partly nonadiabatic operation, on the other hand, can be carried out without any optical potentials, but with the cost of losing a part of the atoms from the trap. Later, Xu \emph{et al.}~\cite{Xu2008} have investigated an alternative pumping cycle in which one of the multipole magnetic fields is replaced by a second homogeneous field but which necessitates the use of an optical trap.

In adiabatic vortex pumping and other methods based on topological phase imprinting, the condensate remains in its instantaneous eigenstate throughout the process, and the final state with a multiquantum vortex is very close to a stationary state. Therefore, the stability properties of stationary multiquantum vortices become essential for determining how large winding numbers can be reached with the pump. Dynamical instabilities and core sizes of giant vortices with large winding numbers were recently studied in Ref.~\cite{Kuopanportti2010}. The maximum strength of the dynamical instability of a $\kappa$-quantum vortex was observed to increase very slowly with $\kappa$. On the other hand, the core size of the vortex, which partly determines the maximum adiabatic pumping speed, was found to increase roughly as $\sqrt{\kappa}$ for large $\kappa$. The results of Ref.~\cite{Kuopanportti2010} suggest that giant vortices with extremely large winding numbers could be created by gradually speeding up the operation of the vortex pump.

Motivated by the promising possibility of realizing isolated vortices with very large winding numbers, we investigate the splitting dynamics and stabilization of such vortex states. We limit the scope of our study to pancake-shaped BECs in nonrotating harmonic traps and study how the giant vortices can be dynamically stabilized with the application of a plug potential in the center of the trap. In order to address the dynamics of vortex splitting, we numerically solve the temporal evolution of the stationary vortex states subjected to small-amplitude perturbations. Based on earlier studies for two- and four-quantum vortices~\cite{splitting1,splitting6}, we expect the splitting to be driven by dynamical instabilities and anticipate the splitting patterns to reflect the symmetry properties of the excitation modes that trigger the splitting processes.

The remainder of this article is organized as follows. In Sec.~\ref{sc:theory}, we present the zero-temperature mean-field theory of the condensate and describe how it is employed in the numerical calculations. Section~\ref{sc:results} presents our results on the stabilization and splitting of giant vortices. In Sec.~\ref{sc:discussion}, we discuss the main results of the work.

\section{\label{sc:theory}Theory and methods}

We restrict our studies to the zero-temperature limit and neglect possible finite-temperature effects. Experiments with dilute BECs can be routinely carried out at temperatures where this approximation is justified~\cite{Leanhardt2003b}. We consider BECs consisting of bosonic atoms of mass $m$ enclosed in nonrotating cylindrically symmetric harmonic traps and assume that the condensates are pancake-shaped, \ie, that the trapping frequencies in the axial and radial directions satisfy $\omega_z \gg \omega_r$. Under these assumptions, the order parameter $\Psi$ of the condensate satisfies the Gross-Pitaevskii (GP) equation, 
\begin{equation}\label{eq:tGPE}
i\hbar\partial_t\Psi(r,\phi,t) = \left[ {\cal H}+ g_{2\mathrm{D}}|\Psi(r,\phi,t)|^2 \right] \Psi(r,\phi,t),
\end{equation}
where the single-particle Hamiltonian is given by 
\begin{equation}\label{eq:Ham}
{\cal H} = -\frac{\hbar^2}{2m}\left(\partial_r^2+\frac{1}{r}\partial_r + \frac{1}{r^2}\partial^2_\phi \right) + V(r).
\end{equation}
The potential $V(r)$ is taken to have the form
\begin{equation}\label{eq:potential}
V(r)=\frac{1}{2}m\omega_r^2 r^2+V_\mathrm{plug}(r),
\end{equation}
where $V_\mathrm{plug}(r)$ denotes the possible optical plug potential. In Eq.~(\ref{eq:tGPE}), the $z$ dependence has been factored out as $\Psi(\br)=\Psi(r,\phi)\exp[-z^2/2a_z^2]/\sqrt[4]{\pi a_z^2}$, where $a_z=\sqrt{\hbar/m\omega_z}$ is the harmonic oscillator length in the axial direction. The order parameter is normalized such that $\|\Psi \|^2 = \int |\Psi(r,\phi)|^2 r\,\ud r \ud\phi = N$, where $N$ is the number of condensed atoms. The effective interaction strength $g_{2\mathrm{D}}$ is related to the vacuum $s$-wave scattering length $a$ by $g_{2\mathrm{D}}=\sqrt{8\pi}\hbar^2 a/m a_z$ and is assumed to be positive. 

Stationary states of the condensates satisfy the time-independent GP equation, which is obtained from Eq.~(\ref{eq:tGPE}) with the replacement $i\hbar\partial_t \longrightarrow \mu$, where $\mu$ is the chemical potential. We choose the stationary states to be axisymmetric vortex states with a quantum number $\kappa$, which implies that the order parameter can be written in the form
\begin{equation}\label{eq:psi}
\Psi(r,\phi) = f(r)e^{i\kappa\phi},
\end{equation}
where $f(r)$ is real and nonnegative. Small-amplitude oscillations about the stationary states can be studied in terms of the Bogoliubov quasiparticles \cite{Pethick2008}. The order parameter is decomposed in the form $\Psi(r,\phi,t)=\exp(-i\mu t/\hbar)\left[\Psi(r,\phi)+\chi(r,\phi,t)\right]$, where the small variation $\chi$ is written as
\begin{equation}\label{eq:oscillation}
\chi=\sum_q \left[ u_q(r)e^{i(\kappa+l_q)\phi-i\omega_q t} + v_q^\ast(r)e^{i(\kappa-l_q)\phi+i\omega_q^\ast t} \right].
\end{equation}
By substituting this decomposition into Eq.~(\ref{eq:tGPE}) and linearizing with respect to $\chi$, we arrive at the Bogoliubov equations
\begin{equation}\label{eq:bogo}
\left( \begin{array}{cc} {\cal{L}}_{\kappa+l_q} & gf^2(r)  \\ -g f^2(r) & -{\cal{L}}_{\kappa-l_q}\end{array} \right) \left( \begin{array}{c} u_q(r) \\ v_q(r) \end{array} \right) = \hbar \omega_q \left( \begin{array}{c} u_q(r) \\ v_q(r) \end{array} \right),
\end{equation}
where 
\begin{equation}
{\cal{L}}_{\kappa}= -\frac{\hbar^2}{2m}\left(\partial_r^2+\frac{1}{r}\partial_r - \frac{\kappa^2}{r^2}\right) + V(r) -\mu + 2g f^2.
\end{equation}
The functions $u_q$ and $v_q$ are the quasiparticle amplitudes corresponding to the eigenfrequency $\omega_q$. Moreover, each excitation is characterized by an integer $l_q$ which determines the angular momentum of the excitation mode with respect to the condensate. Without loss of generality, we will assume that $l_q\geq 0$~\cite{remark1}. The Bogoliubov quasiparticle description is valid only if the oscillatory part $\chi$ has a small norm compared with $\Psi$. 

The eigenfrequency spectrum $\{\omega_q\}$ can be used to analyze the stability of a given stationary state. If the spectrum contains excitations with a positive norm $\int\left[ |u_q|^2-|v_q|^2 \right]r\,\ud r$ but a negative eigenfrequency $\omega_q$, the stationary state is \emph{energetically} unstable. Furthermore, the stationary state is \emph{dynamically} unstable if the quasiparticle spectrum contains at least one eigenfrequency with a positive imaginary part \cite{remark2}. Occupations of such complex-frequency modes initially increase exponentially in time [Eq.~(\ref{eq:oscillation})], and thus small perturbations of a dynamically unstable stationary state typically lead to significant changes in its structure. The complex-frequency modes also quickly drive the system beyond the linear regime of Eq.~(\ref{eq:oscillation}), and hence the complete dynamics must instead be described with Eq.~(\ref{eq:tGPE}).

In the case of a multiquantum vortex, dynamical instability typically signifies that the vortex is unstable against splitting into singly quantized vortices. Moreover, the splitting pattern of the vortex is related to the angular momentum of the complex-frequency mode that induces the splitting~\cite{splitting6}. A mode with the angular-momentum quantum number $l_q$ corresponds to a deformation of the condensate density with $l_q$-fold rotational symmetry. Hence, if the splitting of a multiquantum vortex is driven by an excitation mode with the quantum number $l_q$, the splitting should initially exhibit $l_q$-fold symmetry. In general, complex frequencies have been found to exist only for $l_q \geq 2$ \cite{multi3,multi4,splitting6}.

To address the dynamical stabilization of multiquantum vortices, we study the effect of applying a localized plug potential along the symmetry axis of the trap. For simplicity, we use a step potential to approximate a Gaussian laser beam with a sharp boundary, \ie,
\begin{equation}\label{eq:plug}
V_\mathrm{plug}(r) = A\,\Theta (R_\mathrm{plug}-r),
\end{equation}
where $\Theta$ denotes the Heaviside step function. The dependence on the amplitude $A$ is removed by assuming that the plug is sufficiently strong such that $A\gg \mu$, in which case effectively all particles are removed from the volume occupied by the plug. In this limit of strong pinning, the difference between the steplike and Gaussian-shaped beams becomes negligible~\cite{Tapsa2002}.

Our aim is to investigate the dynamical stabilization and splitting patterns of giant vortices. To this end, we numerically study stationary $\kappa$-quantum vortex states for $\kappa \geq 2$ and their response to slight perturbations. First, we find the stationary states of the form of Eq.~(\ref{eq:psi}) for different values of $\kappa$, $g_{2\mathrm{D}}$, and $R_\mathrm{plug}$ by solving the time-independent GP equation. Utilizing these states, we solve the respective Bogoliubov equations, Eq.~(\ref{eq:bogo}), and search for excitation modes that correspond to eigenfrequencies with large imaginary parts. From the excitation spectra, we also infer the minimum plug widths $R_\mathrm{plug}$ required to dynamically stabilize the vortex states. Second, the obtained stationary states are perturbed slightly by adding either low-amplitude random noise or a small initial population of a complex-frequency quasiparticle mode. The dynamics of the perturbed states are then solved by numerical integration of Eq.~(\ref{eq:tGPE}). In order to relate the dynamics to quasiparticle modes with different angular-momentum quantum numbers $l_q$, we monitor the evolution of different angular-momentum eigenmodes by computing the projections
\begin{equation}\label{eq:projection}
{\cal P}_k \Psi(r,\phi,t) = \frac{1}{2\pi} e^{i k \phi} \int_0^{2\pi}e^{-i k \phi'}\Psi(r,\phi',t)\,\ud \phi',
\end{equation}
where $k$ is an integer. Since the evolution described by Eq.~(\ref{eq:tGPE}) is unitary, we have $\sum_k\|{\cal P}_k \Psi\|^2=N$ at all times.

In the numerical simulations, we use finite difference methods with square $xy$ grids that have roughly 200 points in each dimension. The stationary states are solved using relaxation methods, and Eq.~(\ref{eq:tGPE}) is numerically integrated with the Strang splitting scheme~\cite{Strang1968}. The discretized Bogoliubov equations are solved using the {\footnotesize LAPACK} numerical library implemented in {\footnotesize MATLAB}~\cite{matlab}. In order to have generally applicable results, we measure length in the units of the radial harmonic oscillator length $a_r=\sqrt{\hbar/m\omega_r}$ and time in units of $1/\omega_r$. By normalizing the dimensionless order parameter to unity, the dimensionless interaction strength becomes $\vv{g}=\sqrt{8\pi}N a/a_z$. For example, if we use the values $a = 4.7\,\textrm{nm}$ and $m=1.44 \times 10^{-25}\,\textrm{kg}$ corresponding to ${}^{87}$Rb atoms~\cite{Pethick2008} and set the radial and axial oscillator frequencies to $\omega_r = 2\pi \times 20\,\textrm{Hz}$ and $\omega_z = 50\,\omega_r$~\cite{Clade2009}, we obtain $a_r \approx 2.41\,\mu\mathrm{m}$ and $\vv{g}\approx 0.069\,N$. Therefore, the effective interaction strength for a ${}^{87}$Rb condensate with $N \approx 10^4$ atoms is of the order of $\vv{g} \approx 700$. 

We employ two different types of perturbations in the stationary $\kappa$-quantum vortex states. First, we add a small initial population of a selected complex-frequency excitation mode $q$ such that
\begin{eqnarray}
\Psi(r,\phi,0)&=&f(r)e^{i\kappa\phi}+\eta_\mathrm{m}\Big[ u_q(r)e^{i(\kappa+l_q)\phi} \nonumber \\
&&+ v_q^\ast(r)e^{i(\kappa-l_q)\phi} \Big],
\label{eq:initpsi}\end{eqnarray}
where $0 < \eta_\mathrm{m} \ll 1$ determines the initial mode population and the quasiparticle amplitudes are normalized such that $2\int |u_q(r)|^2 r \,\ud r = 2\int |v_q(r)|^2 r \,\ud r = \int f^2(r) r \,\ud r$. Second, we study the effect of adding random noise in each lattice point $(r_i,\phi_j)$ of the discretized order parameter such that the perturbed state is given by
\begin{equation}\label{eq:noisepsi}
\Psi(r_i,\phi_j,0)=f(r_i)e^{i\kappa\phi_j}+\eta_\mathrm{n}\rho_{i,j}f(r_i),
\end{equation}
where $0<\eta_\mathrm{n} \ll 1$ characterizes the strength of the noise and $\rho_{i,j}$ is a random variable chosen uniformly from the unit disk in the $\mathbb{C}$ plane. Different grid sizes are used to ensure that the effect of the noise is independent of discretization. 
Since the random noise is expected to populate different excitation modes approximately equally, we anticipate the noise-induced splitting to be driven by the dominant complex-frequency mode, \ie, by the quasiparticle mode $q$ with the greatest imaginary part of the eigenfrequency $\omega_q$. Consequently, the splitting is expected to exhibit $l_\mathrm{dom}$-fold rotational symmetry, where $l_\mathrm{dom}$ denotes the angular-momentum quantum number of the dominant complex-frequency mode.

\section{\label{sc:results}Results}

We have solved the time-independent GP equation and Bogoliubov excitation spectra, Eq.~(\ref{eq:bogo}), for different values of the winding number $\kappa$, dimensionless interaction strength $\vv{g}$, and plug width $R_\mathrm{plug}$. The parameters $\kappa$ and $\vv{g}$ were primarily chosen from the experimentally most relevant ranges $2\leq\kappa\leq 50$ and $0\leq\vv{g}\leq 1000$, although in some cases greater values of $\vv{g}$ have also been used. For each $\kappa$, we have determined the smallest width $R_\mathrm{plug}$ of the plug potential, Eq.~(\ref{eq:plug}), which is sufficient to dynamically stabilize the $\kappa$-quantum vortex states in the whole interval $0\leq\vv{g}\leq 1000$. Furthermore, we have computed the temporal evolution of slightly perturbed stationary multiquantum vortices from Eq.~(\ref{eq:tGPE}) for different values of $\kappa$ and $\vv{g}$. The evolution was studied without the plug potential for two types of perturbations, \ie, the addition of either random noise or a particular complex-frequency eigenmode.

\begin{figure}[tb]
\includegraphics[width=220pt]{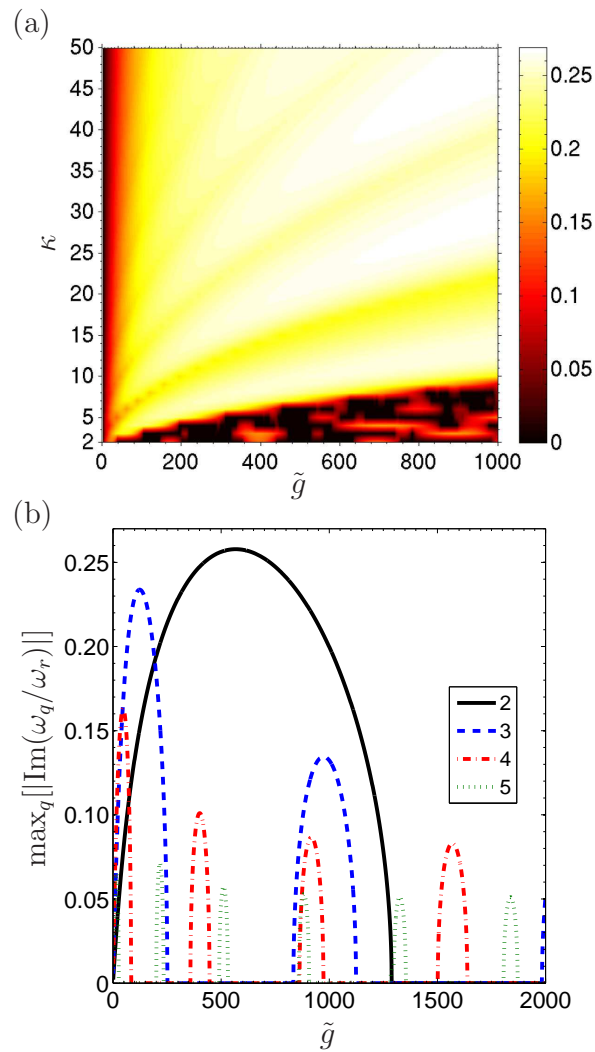}
\caption{\label{fig:maxim}(Color online) (a) Maximum imaginary part of the eigenfrequencies, $\max_q[|\mathrm{Im}(\omega_q/\omega_r)|]$, in the $(\vv{g},\kappa)$ plane. Black areas correspond to vortex states that are dynamically stable. (b) Maximum imaginary part of the eigenfrequencies for the vortex state with $\kappa=10$ as a function of the interaction strength $\vv{g}$ for fixed values of the angular-momentum quantum number $l_q$ given in the inset. For $0 < \vv{g} \leq 2000$, the 10-quantum vortex is dynamically unstable except for three narrow regions within the subinterval $1400 \leq \vv{g} \leq 1950$. The dimensionless interaction strength is given by $\vv{g}=\sqrt{8\pi}N a/a_z$, where $N$ is the particle number, $a$ the $s$-wave scattering length, and $a_z=\sqrt{\hbar/m\omega_z}$ the axial harmonic oscillator length.}
\end{figure}

In the absence of a plug potential, vortex states with large enough winding numbers $\kappa$ were found to be possess dynamical instabilities at nearly all values of $\vv{g}$: vortices with $\kappa \geq 9$ proved to be dynamically unstable in the whole interval $0 < \vv{g}\leq 1000$ [Fig.~\ref{fig:maxim}(a)]. On the other hand, dynamically stable vortex states with $\kappa\geq 9$ were detected at larger interaction strengths [Fig.\ref{fig:maxim}(b)], but no regular behavior for these typically narrow stability regions could be inferred. Within the parameter ranges $2\leq\kappa\leq 50$ and $0 < \vv{g}\leq 1000$, complex eigenfrequencies were found only for angular-momentum quantum numbers $l_q$ satisfying $2 \leq l_q \leq 13$. For fixed values of $\kappa$ and $\vv{g}$, there were typically several dynamical instabilities corresponding to different values of $l_q$ as shown in Fig.~\ref{fig:maxim}(b). The number of different $l_q$ modes supporting complex eigenfrequencies tended to increase with $\kappa$. 

\begin{figure}[tb]
\includegraphics[width=210pt]{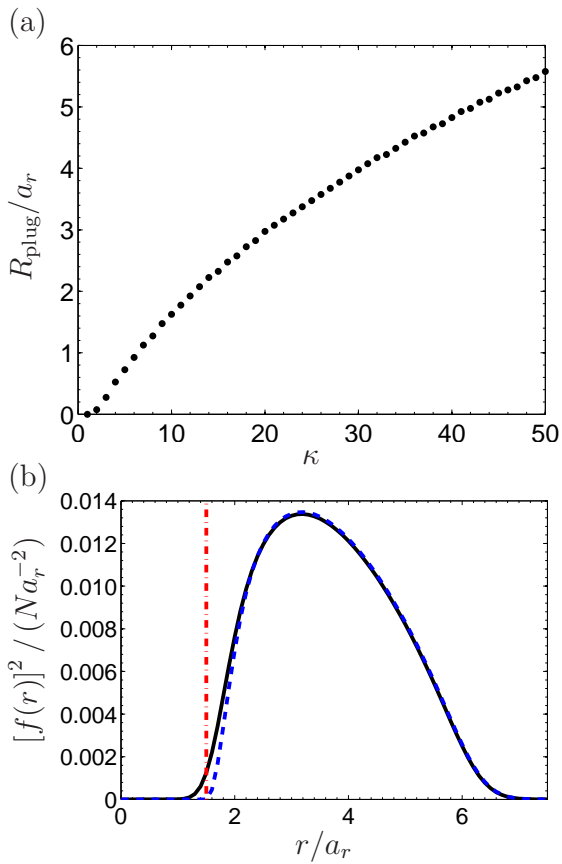}
\caption{\label{fig:plug}(Color online) (a) Minimum plug width $R_\mathrm{plug}$ [Eq.~(\ref{eq:plug})] required to stabilize vortices with a given winding number $\kappa$ in the whole interval $0\leq\vv{g}\leq 1000$. (b) Areal particle densities $\left[f(r)\right]^2$ of the stationary 10-quantum vortex states for $\vv{g}=880$ with (dashed curve) and without (solid curve) a stabilizing plug whose width $R_\mathrm{plug}=1.50\,a_r$ is displayed by the vertical line.}
\end{figure}

The addition of a plug potential of the form of Eq.~(\ref{eq:plug}) was sufficient for dynamically stabilizing the giant vortex states. In Fig.~\ref{fig:plug}(a), we plot, as a function of $\kappa$, the limiting plug width $R_\mathrm{plug}$ at and above which the $\kappa$-quantum vortex is dynamically stable in the whole interval $0\leq\vv{g}\leq 1000$. For simplicity, the results are presented only for the strong steplike beam with only one parameter $R_\mathrm{plug}$, but qualitatively similar behavior was also observed for the linewidth of a Gaussian-shaped beam with a fixed amplitude. The stabilizing value of $R_\mathrm{plug}$ was found to be significantly smaller than the radius of the vortex core, and thus the plug induced only minor changes in the condensate density profile [Fig.~\ref{fig:plug}(b)] or energy. 

To study the different splitting patterns of giant vortices, we have numerically integrated Eq.~(\ref{eq:tGPE}) starting from a slightly perturbed stationary state with given values of $\kappa$ and $\vv{g}$. The optical plug was assumed to be absent, $V_\mathrm{plug}\equiv 0$ [Eq.~(\ref{eq:plug})]. These kind of initial conditions correspond, e.g., to the experimental situation in which the stabilizing plug is switched off adiabatically but much faster than it takes for the vortex to split. We also point out that the addition of a small-amplitude perturbation to a dynamically \emph{stable} vortex state never resulted in splitting, implying that in this case, dynamical instability is a necessary condition for the splitting to occur in the absence of dissipation.

\begin{figure}[tb]
\includegraphics[width=210pt]{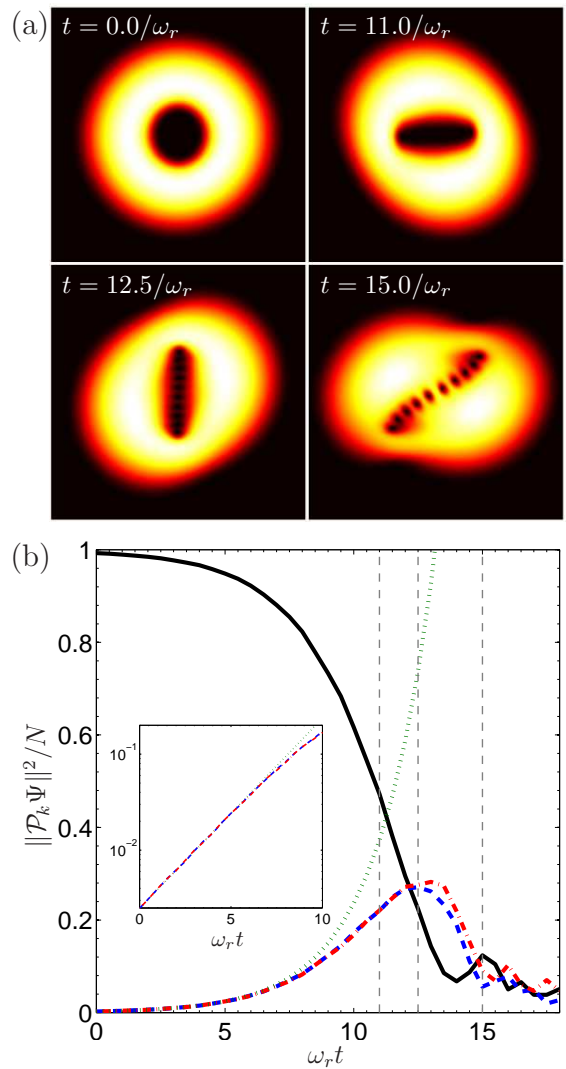}
\caption{\label{fig:k010splitting}(Color online) (a) Density of condensed atoms during the evolution of a 10-quantum vortex state after addition of a complex-frequency mode with quantum number $l_q=2$ and amplitude $\eta_\mathrm{m} = 0.05$. The interaction strength is set to $\vv{g}=880$ and the field of view in each panel is $16\,a_r \times 16\,a_r$. (b) Squared norms of the projections ${\cal P}_k\Psi$ for different values of $k$ [Eq.~(\ref{eq:projection})] as functions of time for the evolution presented in (a). Solid, dashed, and dash-dotted curves correspond to $k=10,12,$ and $8$, respectively, which are the values yielding the largest norms. The dotted curve corresponds to the exponential function $\exp[2\mathrm{Im}(\omega_q)t]=\exp(2\times 0.228\,\omega_r t)$ given by the Bogoliubov equations. The time instants shown in (a) are indicated with vertical lines. The inset shows the same plot for $t\leq 10/\omega_r$ on a semilogarithmic scale (the curve for $k=10$ is not shown).}
\end{figure}

\begin{figure}[tb]
\includegraphics[width=210pt]{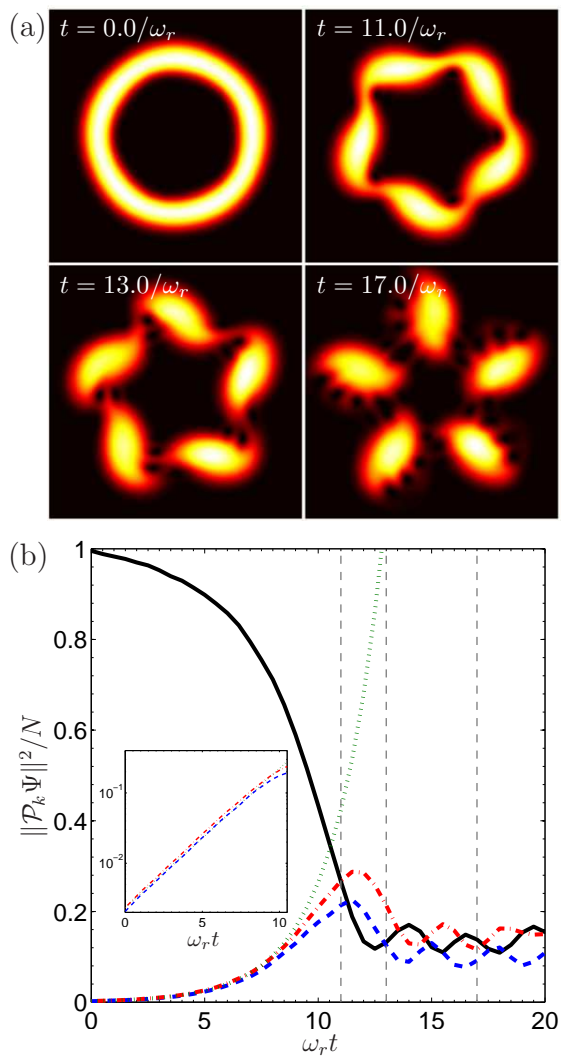}
\caption{\label{fig:k030splitting}(Color online) (a) Particle density of a 30-quantum vortex state with $\vv{g}=200$ after addition of a complex-frequency mode with $l_q=5$ and $\eta_\mathrm{m}=0.05$. The field of view is $18\,a_r \times 18\,a_r$. (b) Squared norms of the projections ${\cal P}_k\Psi$ for different values of $k$ as functions of time for the evolution presented in (a). Solid, dashed, and dash-dotted curves correspond to $k=30,35$, and $25$, respectively. The dotted curve corresponds to the exponential $\exp[2\mathrm{Im}(\omega_q)t]=\exp(2\times 0.236\,\omega_r t)$ given by the Bogoliubov equations. The time instants shown in (a) are indicated with vertical lines. The inset shows the same plot for $t\leq 10/\omega_r$ on a semilogarithmic scale (the curve for $k=30$ is not shown).}
\end{figure}

When a particular complex-frequency mode was used as the initial perturbation, the simulations confirmed the expected correspondence between the splitting pattern and the symmetry of the driving mode: addition of a complex-frequency excitation mode with the angular-momentum quantum number $l_q$ to a stationary multiquantum vortex state resulted in the splitting of the vortex with $l_q$-fold symmetry. Figures~\ref{fig:k010splitting} and \ref{fig:k030splitting} illustrate the typical splitting patterns observed for $l_q=2$ and $l_q=5$, respectively. The populations of the most significant angular-momentum eigenmodes ${\cal P}_k\Psi$ [Eq.~(\ref{eq:projection})] are also shown. The squared norms of the components ${\cal P}_k\Psi$ with $k=\kappa \pm l_q$ initially grow exponentially as predicted by Eq.~(\ref{eq:oscillation}). The first appearance of single-quantum vortices in the density profile is associated with the onset of oscillatory behavior in the populations $\|{\cal P}_k\Psi\|^2$, indicating rapid mixing between a large number of modes with different angular momenta.

\begin{figure}[tb]
\includegraphics[width=220pt]{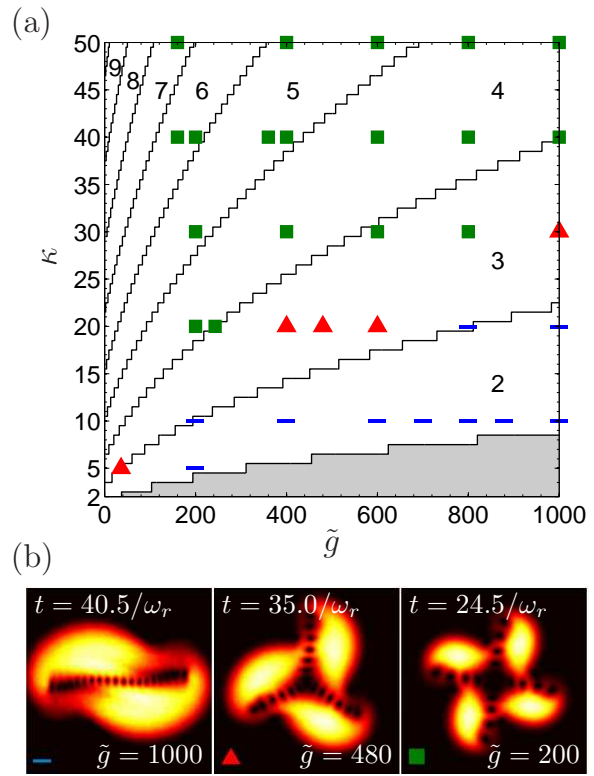}
\caption{\label{fig:splittings}(Color online) Noise-induced splitting of giant vortices. (a) The observed splitting patterns and dominant complex-frequency modes of stationary multiquantum vortices in the $(\vv{g},\kappa)$ plane. The lines separate regions where $l_\mathrm{dom}$, the angular-momentum quantum number of the dominant complex-frequency mode, is constant and has the value indicated. The shaded area corresponds to alternating values of $l_\mathrm{dom}$ and regions of dynamical stability [cf. Fig.~\ref{fig:maxim}(a)]. The bars, triangles, and squares indicate splitting patterns with 2-, 3-, and 4-fold symmetries, respectively, as observed in the numerical integration of the GP equation after addition of random noise of strength $\eta_\mathrm{n}=0.05$ [Eq.~(\ref{eq:noisepsi})]. (b) Density profiles obtained for $\kappa=20$ and different values of $\vv{g}$ illustrating the three types of splitting. The field of view in each panel is $18\,a_r \times 18\,a_r$.}
\end{figure}

When the dynamically unstable vortex states were perturbed with random noise, numerical integration revealed a total of three different splitting patterns, referred to as linear, threefold, and fourfold splitting. The splitting patterns were independent of particular realizations of the noise such that only one type of splitting occurred for given values of $\kappa$ and $\vv{g}$. The results are summarized in Fig.~\ref{fig:splittings}(a), where the observed splitting patterns are compared with the values of the quantum number $l_q$ of the dominant complex-frequency modes. Figure~\ref{fig:splittings}(b) exemplifies the three patterns for $20$-quantum vortices. In linear splitting, the multiquantum vortex disintegrates into a linear chain of single-quantum vortices. In threefold (fourfold) splitting, three (four) sheets of single-quantum vortices emerge symmetrically from the core of the giant vortex toward the surface of the cloud, leaving behind three (four) vortex-free condensate fragments which are separated from each other by the interstitial vortex sheets. Each of the fragments was observed to rotate about the trap center and about its own center. 

The linear, threefold, or fourfold splitting pattern observed after addition of random noise was related to the presence of a complex-frequency mode with the angular-momentum quantum number $l_q=2,3,$ or $4$, respectively, in the excitation spectrum of the corresponding stationary state. Moreover, the noise-induced splitting process was reproduced by using the respective complex-frequency excitation mode as the initial perturbation [Eq.~(\ref{eq:initpsi})]. For example, when the state with $\kappa=10$ and $\vv{g}=880$ was initially perturbed with random noise of strength $\eta_\mathrm{n}=0.05$, the evolution of the condensate density was indistinguishable from the evolution depicted in Fig.~\ref{fig:k010splitting} for the corresponding mode-induced splitting, except that it took until $t=31.0/\omega_r$ before the vortex core had become elongated [panel $t=11.0/\omega_r$ in Fig.~\ref{fig:k010splitting}(a)]. The longer waiting time of noise-induced splitting is explained by the smaller initial population of the driving complex-frequency mode. Finally, we point out from Fig.~\ref{fig:splittings}(a) that the complex-frequency mode accounting for the observed splitting symmetry was not always the one with the largest imaginary part of $\omega_q$.

\begin{figure}[tb]
\includegraphics[width=210pt]{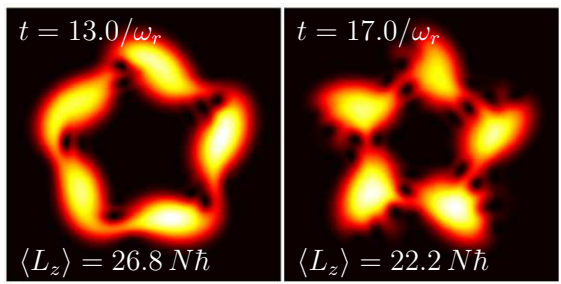}
\caption{\label{fig:k030dissipation}(Color online) Density of condensed atoms during the evolution of a 30-quantum vortex state after addition of a complex-frequency mode with $l_q=5$ and under dissipation of strength $\gamma=0.01$ (see the discussion in the text). Except for the dissipation, the setup is identical to Fig.~\ref{fig:k030splitting}. At $t=0.0/\omega_r$, the expectation value of the axial angular momentum was $\langle L_z \rangle = 30.0\,N\hbar$.}
\end{figure}

We have also studied the effect of dissipation on the splitting dynamics by adding a phenomenological damping term to Eq.~(\ref{eq:tGPE}) through the substitution $t\longrightarrow (1-i\gamma)t$ with $0 < \gamma \ll 1$ and normalizing the order parameter after each time step~\cite{splitting2,Kasamatsu2002,Choi1998,Tsubota2002}. Figure~\ref{fig:k030dissipation} depicts the evolution of a state identical to the one shown in Fig.~\ref{fig:k030splitting}(a) but under dissipation of strength $\gamma = 0.01$. Comparison of Figs.~\ref{fig:k030splitting}(a) and \ref{fig:k030dissipation} shows that dissipation has caused the condensate fragments to shift closer to each other, but otherwise the splitting process has remained unchanged. Similarly, the splitting patterns presented in Fig.~\ref{fig:splittings}(a) were found to remain valid with $\gamma = 0.01$. 

\section{\label{sc:discussion}Discussion}

In this paper, we have investigated the splitting dynamics of multiply quantized vortices in harmonically trapped pancake-shaped BECs for different values of the winding number $\kappa$ and interaction strength $\vv{g}$. The splitting was studied both by directly computing the temporal evolution of the condensate from the time-dependent GP equation and by employing the Bogoliubov equations to predict the essential features of the splitting process. To trigger the splitting, the stationary vortices were perturbed by adding either random noise or a complex-frequency Bogoliubov excitation to the order parameter. In addition, we have studied the dynamical stabilization of the giant vortices by applying a steplike plug potential along the symmetry axis of the trap.

Dynamical instabilities were found to govern the splitting behavior. Here, their existence was a necessary condition for the splitting to occur, and vortices with large winding numbers turned out to be dynamically unstable at nearly all values of the interaction strength $\vv{g}$. On the other hand, a given vortex state could be stabilized by a plug potential of a width somewhat smaller than the size of the vortex core. 

The splitting mechanisms were found to be insensitive to the exact form of the initial perturbations. Even the addition of random noise to a stationary state resulted in a splitting pattern that could be reproduced by perturbing the state with a single complex-frequency mode. Typically, this excitation mode was the one with the maximal imaginary part of the eigenfrequency, although sometimes a complex-frequency mode with a smaller value of the angular-momentum quantum number $l_q$ was excited by the noise instead of the dominant mode [Fig.~\ref{fig:splittings}(a)]. In general, the addition of a complex-frequency mode $q$ resulted in a splitting pattern with $l_q$-fold rotational symmetry. Moreover, the splitting patterns were found to be robust against weak dissipation. 

The splitting of giant vortices with large winding numbers proceeded through the creation of vortex sheets and resulted in separate, vortex-free domains of condensed atoms. The number of these domains was given by the angular-momentum quantum number $l_q$ of the excitation responsible for the splitting. The deformation of the giant vortex core and the eventual fragmentation of the surrounding condensate region were accompanied with large oscillations in the populations of different angular-momentum eigenmodes, suggesting that the formation of vortex sheets is a strongly nonlinear phenomenon.

Our results indicate that a giant vortex does not split by melting into a chaotic liquid of singly quantized vortices. Instead, the long-time dynamics preserves the rotational symmetry of the excitation mode which causes the splitting. Therefore, the vortex pump with the simplest pumping cycle may not be an efficient way to realize a strongly correlated vortex liquid phase~\cite{Cooper2001} as envisioned in Ref.~\cite{pumppu}. 

In the future, it would be interesting to study the splitting in a fully three-dimensional geometry to find out if additional effects such as intertwining of the vortex cores~\cite{splitting2,splitting4,splitting5} would lead to more chaotic dynamics. For example, the intertwining might cause the vortex-free domains to intertwine along the axial direction~\cite{Eltsov2006}. 

The experimental creation of a nearly pure giant vortex state in a pancake-shaped BEC, \eg, using the vortex pump, and the consequent observation of its splitting would provide a test for the validity of the Bogoliubov quasiparticle description. The splitting of the condensate into distinct fragments is detectable with current imaging techniques, and would allow for the clear identification of the angular-momentum quantum number of the driving excitation. In addition, it would be interesting to determine if the phase coherence between the separate condensate fragments could be lost due to the rapid phase variations caused by the interstitial vortices.

\begin{acknowledgments}
The authors acknowledge the Academy of Finland, the V\"ais\"al\"a foundation, and the Emil Aaltonen foundation for financial support. J.~A.~M. Huhtam\"aki, E. Lundh, and V.~Pietil\"a are appreciated for helpful discussions.
\end{acknowledgments}

\bibliography{splitting}

\begin{thebibliography}{47}
\bibitem[{com()}]{remark1}
\bibinfo{note}{Equations~(\ref{eq:oscillation}) and (\ref{eq:bogo}) imply that for each eigensolution $(u_q,v_q;\omega_q,l_q)$, there is another solution $(v_q^\ast,u_q^\ast;-\omega_q^\ast,-l_q)$ which corresponds to the same variation of $\Psi$.}
\bibitem[{com()}]{remark2}
\bibinfo{note}{In fact, it is sufficient to require only that the spectrum contains an eigenfrequency with a nonzero imaginary part since if $(u_q,v_q;\omega_q,l_q)$ is a solution of Eq.~(\ref{eq:bogo}), then $(u_q^\ast,v_q^\ast;\omega_q^\ast,l_q)$ is also a solution.}
\end{thebibliography}
\end{document}